# Measurement of the Blackbody Radiation Shift of the $^{133}$Cs Hyperfine Transition in an Atomic Fountain

Filippo Levi , Davide Calonico, Luca Lorini, Salvatore Micalizio and Aldo Godone

*Istituto Elettrotecnico Nazionale "Galileo Ferraris", Strada delle Cacce 91, 10135 Torino, Italy*

We used a Cs atomic fountain frequency standard to measure the Stark shift of the ground state hyperfine transition frequency in cesium (9.2 GHz) due to the electric field of the blackbody radiation. The measured relative shift at 300 K is $(-1.43\pm0.11)\times10^{-14}$ and agrees with our theoretical evaluation $(-1.49\pm0.07)\times10^{-14}$. This value differs from the currently accepted one $(-1.69\pm0.04)\times10^{-14}$. The difference has a significant implication on the accuracy of frequency standards, in clocks comparison, and in a variety of high precision physics tests such as the time stability of fundamental constants.

Laser cooled cesium fountains are presently the most accurate frequency standards as they allow to reach the $1\times10^{-15}$ accuracy level [1-4]. These capabilities offer great opportunities in a variety of precision measurements, such as frequency metrology and realization of the International Atomic Time (TAI) [5], high accuracy spectroscopy [6], laboratory tests of fundamental constant time stability [7] and therefore validation of general relativity and string theory [8].
In a fountain, the reference Cs hyperfine transition frequency in the $6^2S_{1/2}$ level is perturbed by four main effects: the density shift caused by cold Cs atomic collisions [9], the Zeeman shift induced by the quantization magnetic field [10], the red shift due to the Earth gravitational potential (clearly, this effect is relevant only for frequency comparison among clocks at different height on the geoid) [10] and finally the AC Stark shift induced by the Black Body Radiation (BBR) of the environment (the AC Zeeman shift is three order of magnitude lower) [11]. These shifts have to be corrected in order to match the definition of the second in the International System of Units (SI) [12] and therefore they have to be known at the best accuracy level.

The AC Stark shift induced by the blackbody radiation (BBR) was pointed out as a relevant bias for Cs atomic clocks in the early '80s [13] with the following relation:

$$\Delta\nu/\nu_0 = \beta(T/300)^4\left[1+\varepsilon(T/300)^2\right] \quad (1)$$

where T is the radiation temperature of the environment expressed in kelvin, and $\nu_0$=9192631770 Hz. The temperature dependence in (1) is quite general since it derives directly from the Planck and the Stefan-Boltzmann radiation laws. The coefficients β and ε depends on the atom and on the considered transition; up to now the theoretical accepted value of β and ε for $^{133}$Cs are $\beta=(-1.69\pm0.04)\times10^{-14}$ and $\varepsilon=1.4\times10^{-2}$.
Experimental determinations of β follow two different methods. The first one is indirect as it derives the AC Stark shift from a measurement of the DC Stark shift produced by a static electric field. This method has been used by [14,15] and by [16] where respectively an atomic beam apparatus or a Cs fountain were used. Each of them rely on the application of an electric field of several kV/cm in the atoms free flight region between the two separated interaction regions of a Ramsey scheme. The value obtained for β are consistent among them and with the value reported above.
The second one is a direct method: a Cs frequency standard is operated at different temperatures to evaluate the BBR shift. Up to now, the only direct measurement is reported in [17] by use of a Cs thermal beam standard and the value obtained is $\beta = (-1.66\pm0.20)\times10^{-14}$, in agreement at 1σ with the value reported in the indirect experiments.

In this Letter we report the first direct measurement of the BBR shift using a Cs fountain: the measured value is $\beta = (-1.43\pm0.11)\times10^{-14}$ in agreement at the 1σ level with the direct measurement reported in [17], but smaller of about 15% than the values deduced by using indirect measurements [14-16]. We also report the result of a new theoretical evaluation of the BBR shift we have performed, which agrees with our experimental value.
This result has significant implications on the field of precision frequency measurements, as it involves the first element of the frequency metrology chain, the primary standards. In fact, at room temperature the difference in the BBR shift evaluation with the old and the new value is about $3\times10^{-15}$, nearly three times the accuracy declared for the standards themselves. Moreover, another striking difference arises in the comparison of two standards operating at different temperatures: for example, two Cs standards working at 300 K and 340 K respectively will show a $2\times10^{-15}$ difference in the frequency comparison while using one or the other value of β.

Our atomic fountain (IEN-CSF1) was described in details in [4] and here we recall only the operation principle and some details relevant for the present experiment.
Up to $5\times10^7$ $^{133}$Cs atoms are collected and cooled down to ~1μK in a magneto-optical trap in 0.2 s; then, after a free expansion time that reduces the atomic density, the cesium molasses is vertically launched in a ballistic flight. Atoms in the hyperfine ground state |F=3, $m_F$=0⟩ are selected by a combination of microwave and laser pulses and then undergo the Ramsey interaction.

The Ramsey interaction is realized by a temperature-tuned microwave cavity placed along the flight trajectory where the atoms pass twice, first on the way up and then on the way down; exiting the cavity we detect separately the atoms in |F=3, $m_F$=0⟩ and in |F=4, $m_F$=0⟩ to obtain the transition probability.

The cavity and the drift tube are realized in OFHC copper and are also the side of the vacuum chamber. The temperature of the fountain structure can be varied by means of heaters. The state selection and the Ramsey interaction take place inside a magnetically shielded area where a solenoid generates a quantization magnetic field (C-field) along the z axes of nominal value $B_0 \sim 10^{-7}$ T.

The microwave frequency is synthesized from a low phase noise 5 MHz BVA quartz oscillator, which is phase locked to a hydrogen maser. A scheme of the fountain structure is shown in Figure 1.

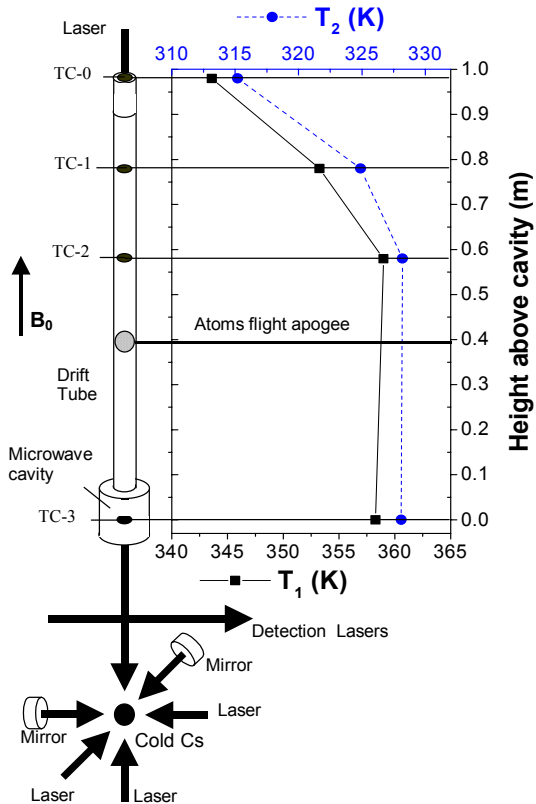

FIG. 1. Schematic structure of the fountain. The temperature sensor positions are reported, together with the temperature profiles along the structure at T1 (358 K) and T2 (328 K). TC-0, TC-1, TC-2 and TC-3 indicate the four temperature sensors.

In our experiment we measure the frequency of the fountain with respect to that of the hydrogen maser (HM) used as reference: the fountain alternatively works at two different temperatures $T_1$ and $T_2$ and β is evaluated from the difference of the two measured frequencies. Indeed, since every measurement has about $1 \times 10^{-15}$ uncertainty, a difference between $T_1$ and $T_2$ of at least 30 K is required around 300 K to obtain a frequency difference of about $10^{-14}$ and an uncertainty below $10^{-15}$ for β.

The HM frequency (drift removed) is very stable on the medium and long term ($3 \times 10^{-16}$ over one week) and allows to compare the fountain frequency measurements at different temperatures spreading over few weeks. A very careful estimation of the HM drift is a key point of the experiment. For this purpose, each measurement of β is composed of four subsets of fountain frequency evaluations, alternated at $T_1$ and $T_2$ and moreover, every single subset lasts at least five days.

The sets of data are then processed by a multilinear least squares technique, where two parallel lines fit the data at $T_1$ and $T_2$. The parallel lines model uses only three parameters, the maser drift and the axes intercepts. The frequency difference between the atomic fountain in $T_1$ and $T_2$ is given directly by the difference of the ordinate intercepts of the two lines as evaluated by the fit, as shown in Figure 2 where the measured data in one run of the experiment are reported.

In order to check the evaluation of the drift obtained by the fit, we have compared by means of a two way satellite time and frequency transfer (TWSTFT) [18] our maser with the remote time scales UTC(NIST) and UTC(NPL) of the National Institute of Standards and Technology (USA) and of the National Physical Laboratory (UK) respectively. These evaluations of the maser drift $d$ agree with our local measurement $d = (-9.7 \pm 0.4) \times 10^{-16}$/day within $1 \times 10^{-16}$/day.

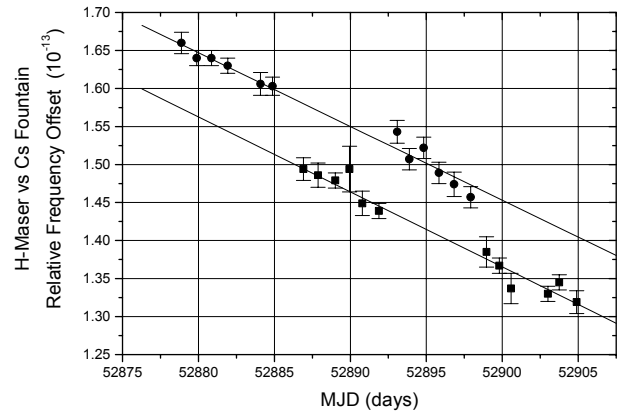

FIG. 2. Measurements of the relative frequency difference between the fountain and the hydrogen maser during one run of the experiment versus the Modified Julian Date (MJD). ● T=328 K, ■ T=358 K, ─ bi-linear fit.

In IEN CS-F1 the Ramsey cavity is temperature-tuned to the Cs resonance frequency; its working point is $T_0$=343 K and it is not possible to change the temperature of the drift region without changing the cavity temperature. The experiment requires to operate the cavity out of resonance and then it is necessary to increase the microwave power. As the loaded quality factor of the cavity is $Q_L \sim 20000$ and the frequency/temperature sensitivity is 150 kHz/K, a 15 K off resonance condition requires an increase of 20 dB of the microwave power in order to maintain the optimum level for the Ramsey spectroscopy (π/2 microwave pulse). This power increase could result in undesirable frequency shifts of the clock transition due to the presence of possible microwave leakage in the apparatus. We have not detected any frequency shift at the $1\times10^{-15}$ level in differential measurements on IEN-CsF1 when operating, at resonance, alternatively with π/2 and 5π/2 microwave pulses.

Furthermly, a key point of the measurement is to choose two different temperatures $T_1$ and $T_2$ (around 328 K and 358 K respectively) equally spaced with respect to the usual working temperature $T_0$.

In this way, the differential measurement rejects all the unwanted microwave shifts, being the microwave power unchanged at $T_1$ and $T_2$.

Another significant improvement of our measurement with respect to the previous one [17] is that the Ramsey interaction in a fountain is performed with a single cavity, so no end to end phase shift effect exists. The effect of the cavity pulling, linear and quadratic [2], is evaluated to be $<1\times10^{-16}$.

Moreover, differential determination of β allows to reach a type-B uncertainty of $5\times10^{-16}$ in spite of the fact that the IEN-CSF1 accuracy is currently evaluated to be $2\times10^{-15}$. In fact, IEN-CSF1 accuracy is limited by the density shift evaluation, but since the average density is kept constant at 10 % during each run of the BBR experiment, the atomic density shift is cancelled at the $4\times10^{-16}$ level.

Another temperature dependent effect in the experiment is the Zeeman shift: in fact, the pitch of the solenoid that provides the C-field can change, resulting in a different magnetic shift on the transition at $T_1$ and $T_2$, as well as it may happen if thermoelectric currents are present in the structure. We have done two C-field maps at $T_1$ and $T_2$ to evaluate the Zeeman shift in the two situations and we have measured a maximum frequency difference $<1\times10^{-16}$.

The last possible temperature dependent effect which could affect the fountain frequency is a change in the copper surface outgasing (hydrogen); we have calculated for this effect a conservative uncertainty of $1\times10^{-16}$.

The fountain structure, even if not originally designed for this purpose, is a very good black body radiator. The cavity and the drift tube (see Figure 1) are both realized with the same material (OFHC copper) whose high thermal conductivity ensures a good enough temperature homogeneity. Seen from the inside, from the atoms point of view, the fountain structure is a cavity at a given temperature with two very small holes at the ends. Furthermore the window at the top end is made of BK7 glass and is extremely dark in the far-infrared where the BBR spectrum reaches its maximum (~17 THz).

The temperature is measured by four type T thermocouples along the interaction region, one on the cavity body, one close to the atom apogee, one 20 cm above the apogee and one at the fountain top on the upper window.

Three heaters set the temperature of the environment, two inside the interaction region (one on the cavity and one on the drift tube) and one on the upper window. The heaters consist of a constantan wire fed by an alternated current at 30 kHz to avoid the generation of a magnetic field that could perturb the atomic sample. A servo system keeps the temperature constant within 0.04 K. The typical temperature profiles of the structure at $T_1$ and $T_2$ are also reported in Figure 1. We observe that the atoms, in their parabolic flight, spend most of the time where the radiation temperature is more homogeneous.

The skin effect of the copper structure reduces strongly the penetration inside the interaction region of the RF current generated by the heater. No AC Zeeman effect was observed at the $8\times10^{-16}$ level, uncertainty that is further reduced in the differential measurements done at the level of $3\times10^{-16}$.

To calculate the effective blackbody temperature seen by the atoms we interpolate the measured temperatures with a polygonal curve and then we calculate the average radiation temperature experimented by the atoms at a given position (integrated over the solid angle); in this way it is possible to calculate also the effect of the two "holes" in the blackbody radiator, the upper window and the hole in the microwave cavity. The values obtained at different elevations inside the fountain structure are then used to calculate the time averaged radiation temperature seen by the atoms along their ballistic flight.

The effect of both the upper and the lower holes is very small, and the variation of radiation temperature for one degree of variation of the hole temperature is $3\times10^{-4}$ K/K in the considered temperature range.

The surface emissivity of the oxided copper reported in literature varies from 0.5 to 0.88; we have calculated that in our geometrical configuration it does not affect the quality of the BBR radiator at the accuracy level of the experiment.

The radiation temperature $T_1$ and $T_2$ and the frequency difference measured are used then to measure the β coefficient according to the formula (1). At the resolution level of our experiment and with only two different temperature points it is not possible to check the value of the parameter ε; we assume for it the theoretical value ε $=1.4\times10^{-2}$ given in [13].

We have measured β in three different runs; one of them is shown in Figure 2, while all the measurements performed are reported in Table I. The final weighted average value is

$$\beta = (-1.43 \pm 0.11)\times10^{-14}$$

TABLE I. Measured values of the BBR shift coefficient β and related type A and type B uncertainties.

|  | β ($10^{-14}$) | $u_A$ ($10^{-14}$) | $u_B$ ($10^{-14}$) |
|---|---|---|---|
| Run 1 | -1.50 | 0.19 | 0.07 |
| Run 2 | -1.47 | 0.17 | 0.07 |
| Run 3 | -1.40 | 0.10 | 0.07 |
| Average | -1.43 | 0.11 | |

The final uncertainty is the quadratic sum of the statistical (type A) and of the systematic (type B)

uncertainties; the type B uncertainty is evaluated as the maximum fluctuation of the fountain biases from measurement to measurement, plus the uncertainty in the knowledge of the relevant parameters (e.g. radiation temperature) as reported in Table II.

TABLE II. Typical type B uncertainty budget for a single measurement run of β.

| Effects | Uncertainty |
|---|---|
| Radiation temperature | $4 \times 10^{-16}$ |
| Density fluctuations | $4 \times 10^{-16}$ |
| AC and DC Zeeman effect | $5 \times 10^{-16}$ |
| Microwave leakage | $<1 \times 10^{-16}$ |
| Cavity Pulling | $<1 \times 10^{-16}$ |
| Background gas | $\leq 1 \times 10^{-16}$ |
| Total | $7 \times 10^{-16}$ |

Our experimental measurement of the BBR shift is compatible with the previous direct measurement [16] with an improved accuracy of a factor two, but is significantly different from the usually accepted value. We have also performed a new theoretical evaluation of the BBR shift taking into account i) the most recent physical data on the Cs atom as regards the electric dipole moments [19 and references therein] and the frequencies of the D1 and the D2 lines [20], which give the highest contribution to β; ii) a basis of modified eigenfunctions which accounts for the hyperfine interaction of the S and P states with other states of the same quantum numbers F and $m_F$ as pointed in [21]. The result of our analysis is: $\beta=(-1.49\pm0.07)\times10^{-14}$ and $\varepsilon=1.4\times10^{-2}$ and will be reported with more details in a forthcoming paper.

The agreement between our theoretical and experimental values is widely satisfactory. Furthermore we notice that most of the recent data reported in the literature on the polarizability of the $6^2S_{1/2}$ Cs ground state [22], obtained with different experimental techniques, are lower than those obtained with the DC Stark effect measurements. Since it is commonly accepted that the BBR shift should scale with polarizability, these recent data suggest that the coefficient β could be lower than that accepted up to now, in agreement with our results.

It is possible that the DC Stark effect measurements [14-16], in agreement among them, were affected by some systematic effect due to the fact that the DC electric field is not applied continuously during the free flight phase of the Ramsey interaction but as a fast step.

In conclusion we have reported a direct measurement of the BBR shift of the $^{133}$Cs hyperfine ground state transition, firstly performed in an atomic fountain with an uncertainty of $1\times10^{-15}$. Our experimental value agrees both with a theoretical re-evaluation of the BBR shift and with recent data on the $^{133}$Cs atomic polarizability and differs by about $3\times10^{-15}$ from the commonly accepted value; this could require to reconsider the accuracy of the Cs atomic frequency standards used in TAI and in some fundamental physical tests.

We wish to thank sincerely Dr. P. Tavella for her help and fruitful discussion during the data analysis and Dr. M.L. Rastello for fruitful discussions on the Blackbody radiators.